\documentclass[cup5b]{caps}
\usepackage{graphicx}
\usepackage{amssymb}
\usepackage{ociwsymp1}   %Use this for invited papers.
\usepackage{natbib}

\newcommand{\sun}{\odot}

\newcommand {\kms}{\ensuremath{\mathrm{km\,s}^{-1}}}

\bibpunct[,]{(}{)}{;}{a}{}{,} % To get rid of comma between autors' names and years$
\newcommand{\trh}{\ensuremath{t_\mathrm{rh}}}
\newcommand{\tcoll}{\ensuremath{t_\mathrm{coll}}}
\newcommand{\trel}{\ensuremath{t_\mathrm{rel}}}
\newcommand{\tcc}{\ensuremath{t_\mathrm{cc}}}
\newcommand{\tsegr}{\ensuremath{t_\mathrm{segr}}}
\newcommand{\tstar}{\ensuremath{\tau_\ast}}

\newcommand{\remove}[1]{} % To remove text from the compiled document but leave it in the source

\HeadText{F.\ A.\ Rasio, M.\ Freitag, and M.\ A.\ G\"urkan}  

\def\plotone#1{\centering \leavevmode
\includegraphics[width=.95\columnwidth]{#1}}

\begin{document}

\pagenumbering{arabic}

% Corrected Atakan's name.  -F. Rasio
\author[]{FREDERIC A.\ RASIO$^{1}$, MARC FREITAG$^{1,2}$, and M.\ ATAKAN G\"URKAN$^{1}$
\\
(1) Department of Physics and Astronomy, Northwestern University, Evanston, 
IL 60208, USA \\
(2) Astronomisches Rechen-Institut, M\"onchhofstrasse 12-14, D-69120 
Heidelberg, Germany}

\chapter{Formation of Massive Black Holes \\ in Dense Star Clusters}

\begin{abstract}
We review possible dynamical formation processes for central massive black 
holes in dense star clusters. We focus on the early dynamical evolution of 
young clusters containing a few thousand to a few million stars. One natural 
formation path for a central seed black hole in these systems involves the 
development of the Spitzer instability, through which the most massive 
stars can drive the cluster to core collapse in a very short time. The sudden 
increase in the core density then leads to a runaway collision process and the 
formation of a very massive merger remnant, which must then collapse to a 
black hole. Alternatively, if the most massive stars end their lives before 
core collapse, a central cluster of stellar-mass black holes is formed.
This cluster will likely evaporate before reaching the highly relativistic 
state necessary to drive a runaway merger process through gravitational 
radiation, thereby avoiding the formation of a central massive black hole. We 
summarize the conditions under which these different paths will be followed, 
and present the results of recent numerical simulations demonstrating the 
process of rapid core collapse and runaway collisions between massive stars.
\end{abstract}

\section{Introduction}

The main focus of this chapter is on the formation of massive black holes 
(BHs) through stellar-dynamical processes in young star clusters. Here by 
``massive'' we mean BHs with masses in the range $\sim 10^2-10^4\,M_\odot$, 
which could be ``intermediate mass'' BHs in small systems such as globular 
clusters (van der Marel, this volume) or ``seed'' BHs in 
larger systems such as proto-galactic nuclei. The later growth of seed BHs by 
gas accretion or stellar captures to form supermassive BHs in galactic nuclei 
is discussed by Blandford (this volume).  The early dynamical evolution of 
dense star clusters on a time scale $t\lesssim 10^7\,{\rm yr}$ is completely 
dominated by the most massive stars that were formed in the cluster. This 
early phase of the evolution is therefore very sensitive to the stellar 
initial mass function (IMF), particularly at the high-mass end (see Clarke, 
this volume). One possibility, which we will discuss in some detail here, is 
that the massive stars could drive the cluster to core collapse before 
evolving and undergoing supernova explosions. Successive collisions and 
mergers of these massive stars during core collapse can then lead to a 
runaway process and the rapid formation of a very massive object containing 
the entire mass of the collapsing cluster core. Although the fate of such a 
massive merger remnant is rather uncertain, direct ``monolithic'' collapse to 
a BH with little or no mass loss is a likely outcome, at least for 
sufficiently low metallicities \citep{HFWLH02}. 

Alternatively, if the most massive stars in the system evolve and produce 
supernovae before the onset of this runaway collision process, the collapse of 
the cluster core will be reversed by the sudden mass loss, and a cluster
of stellar-mass BHs will be formed. The final fate of this cluster is also 
rather uncertain.  For small systems such as globular clusters, complete 
evaporation is likely (with all the stellar-mass BHs ejected from the cluster 
through 3-body and 4-body interactions in the dense core). This is expected 
theoretically on the basis of simple qualitative arguments (Kulkarni, Hut, 
\& McMillan 1993; Sigurdsson \& Hernquist 1993)
and has been demonstrated recently by direct $N$-body 
simulations \citep{PZMcM00}.  However, for larger systems such as 
proto-galactic nuclei, contraction of the cluster to a highly relativistic 
state could again lead to successive mergers (driven by gravitational 
radiation) and the formation of a single massive BH \citep{QS89}.

Because of the great complexity and variety of dynamical processes
involved, questions related to BH formation in dense star clusters are
best studied using numerical simulations. On today's computers, this
can be done with direct $N$-body simulations for $N\approx 10^3 - 10^5$
stars \cite[see, e.g.,][]{Aarseth99} or with Monte Carlo (MC) techniques
(see Sec.~1.2) for up to $N\approx 10^7$ stars. Observationally, this large
range covers a variety of well-studied young star clusters, from
``young populous clusters'' (e.g., Arches, R136) to ``super star
clusters'' \citep{Whitmore00}. However, realistic, star-by-star
simulations of larger systems, on the scales of entire galactic
nuclei, are not yet possible. For these systems, one must rely on more
qualitative analyses based on extrapolations from numerical results
for smaller $N$.

The role of stellar collisions in 
dense galactic nuclei was first consider to explain the quasar/active galactic 
nucleus phenomenon (e.g., Spitzer \& Saslaw 1966). \citet{Colgate67} pointed out
that, when collisions first set in a stellar cluster evolving from
``reasonable'' initial conditions, they are unlikely to be disruptive
so that runaway mergers should lead to the formation of massive
stars. He estimated that growth would saturate at $\sim 50\,M_\sun$
because small stars could fly across a  massive star without being
stopped. In ``particle-in-a-box''-type simulations using
a semi-analytical model for the outcome of individual collisions,
\citet{Sanders70b} found clear runaway growth up to a few hundred $M_\sun$. 
No proper account of the stellar dynamics was included, however; the
cluster was treated as a homogeneous sphere of constant mass (the gas
ejected in collisions being recycled into stars) contracting through
collisional energy loss. 

Stellar dynamics must be playing an important role, however, especially for 
massive stars, which can be affected significantly by mass segregation.
Indeed, massive stars of mass $m$ undergo mass segregation and concentrate 
into the cluster core on a time scale 
$\tsegr \simeq (\langle m \rangle/ m)\, \trh $,
where $\langle m \rangle$ is the average stellar mass and the overall 
relaxation time (at the half-mass radius $r_{\rm h}$) for a cluster of total 
mass $M$ is given by
$$ \trh \simeq 10^8\,{\rm yr}\,\left(r_{\rm h}\over 1\,{\rm pc}\right)^{3/2}
\left(M \over 10^6\,M_\odot \right)^{1/2} \left( \langle m \rangle \over 
1\,M_\odot\right)^{-1}. $$
It is clear that the most massive stars in a dense cluster can undergo mass 
segregation on a time scale much shorter then their stellar evolution time.
When they eventually dominate the density in the cluster core, these massive 
stars will then {\em decouple dynamically\/} from the rest of the cluster (go 
out of thermal equilibrium, evolving {\em away\/} from energy equipartition) 
and evolve very quickly to core collapse. This process is
often referred to as Spitzer's ``mass-segregation instability.''

The possibility of a ``mass-segregation instability'' in simple two-component 
systems (clusters containing only two kinds of stars, one much more massive 
than the other) was first predicted by \citet{Spitzer69}, and the first 
dynamical simulations revealing mass segregation at work were performed in the 
70's \citep{SH71b,SS75b}. The physics of this instability is now very well 
understood theoretically (Watters, Joshi, \& Rasio 2000).  In a remarkably 
prescient paper, \citet{Vishniac78} showed correctly for the first time that 
this instability must affect the early dynamical evolution of any star cluster 
born with a reasonable ``Salpeter-like'' IMF. He concluded that, as a result, 
``most globular
clusters may undergo core collapse at an early time in the evolution of 
the universe. This is clearly significant as a possible mechanism for creating 
collapsed bodies in the center of globular clusters.''\footnote{Historical note 
contributed by G.~Burbidge: ``When we first were trying to understand what was 
responsible for violent events, originally in radio sources, in the early 
1960s we (Hoyle, Fowler and the Burbidges') discussed the problem of 
gravitational collapse of massive objects (superstars) after Hoyle and Fowler 
had argued that the mechanism I had proposed in 1961, chain reactions of 
supernovae, would not work.  In this early period Hoyle and Fowler wrote 
several papers on the collapse, and in 1964 $HFB^2$  published a paper in ApJ 
discussing all of the ramifications. We did not know how to form the beast 
--- the work of Colgate and Spitzer showed how it would work if the star 
density was high enough, but the densities required were far greater than any 
that could reasonably exist in the centers of galaxies.  But we were convinced 
that such phenomena must exist in the centers, and Hoyle was quite angry when 
much later in 1969 Lynden-Bell was given credit for the idea after a press 
conference.  By then Hoyle and I had become convinced that the usual mechanism 
of accretion would not give enough energy because the efficiency of the 
process in terms of what {\em we see\/} must be very low, and much below 
10\%.  Thus we turned to new and fundamental ideas concerning creation in 
galaxy centers which can only take place in regions of very strong 
gravitational fields, i.e., next to classical BHs.  This is my 
position today.  Fred and I worked on it until his death.  When Vishniac's 
paper appeared in 1978, Fred and I discussed it and realized that mass 
segregation might allow a BH to form rapidly.  But this was in the 
context of total masses of  $10^5 - 10^6\,M_\odot$  --- globular clusters --- 
much less than is required for massive BHs in nuclei. Still we talked 
about doing more work on it.  It was not pursued, probably because we were 
neither in a position to do it ourselves; I was just moving to Kitt Peak as 
director, and Fred had left Cambridge and was living up in Cumberland. But we 
both thought it was an important step forward.''} In their classic paper the 
same year, \citet{BR78} also mentioned the combination of mass segregation and 
runaway merging as ``one of the quickest routes to the formation of a 
massive object in a dense stellar system.'' 

However, the first detailed dynamical simulations of dense star clusters 
including stellar collisions and mergers considered clusters where all stars 
are initially identical (Lee 1987; Quinlan \& Shapiro 1990, hereafter QS90).
These Fokker-Planck (FP) simulations
also included the dynamical formation of binaries through 3-body
interactions and their subsequent hardening (and ejection) as a
central source of energy capable of reversing core collapse and
turning off collisions in clusters with a relatively low number of
stars. Furthermore, in these simulations, collisions themselves can stop
collapse when collisionally produced massive stars lose mass in a supernova
explosion at the end of their life. The results of these early simulations 
suggested that runaway collisions would occur provided that $\trh < 10^8\,$yr 
(to beat stellar evolution) and $N \ge 3\times 10^6$ (to avoid binary 
heating). QS90 stressed that, as a result of mass segregation, the 
rise in the central velocity during collapse is only moderate and collisions
do not become disruptive.  Although not very realistic, these early studies 
made plausible the idea that successive collisions and mergers of 
main-sequence stars could lead to the formation of a 
$\sim 10^2 - 10^3\, M_\odot$ object. 

More recently, through direct $N$-body simulations of clusters containing
2000 to 65,000 stars, \citet{PZMcM02} showed that, in such
low-$N$ systems, dynamically formed binaries, far from preventing
collisions (by heating the cluster and reversing collapse), actually 
{\em encourage\/} them by
increasing the effective cross section. In these small systems, 
once the few massive stars have
segregated to the center, one of them will repeatedly form a binary
with another star and later collide with its companion when an
interaction with a third star increases the binary's eccentricity. The
growth of this star is ultimately stopped by stellar evolution, or by
the dissolution of the cluster in the tidal field of the parent galaxy. Given 
the small number of stars in these simulations, the maximum mass of the 
collision product is only $\sim 200\,M_\sun$ when mass loss from stellar winds 
is negligible.  The ``pistol star'' in the Galactic Center Quintuplet cluster 
\citep{FNMMcLGGL98} may provide an example of a directly observable runaway 
collision product of this type in a small, young star cluster.

Work in progress by the authors is now attempting for the first time to study 
numerically these processes for much larger star clusters, containing up 
to $N\approx 10^7$ stars, and resolving in detail the core collapse and runaway 
collisions. Section 1.2 provides a summary of the numerical methods that we are 
using, based on a MC technique for collisional stellar dynamics.
Section 1.3 presents a few of our initial results. Our main conclusions can be 
summarized as follows.

Our numerical simulations show that, in the absence of stellar evolution, the 
core collapse time in a dense star cluster is always given by
$$ \tcc \simeq 0.1\, \trh, $$ for any ``reasonable'' IMF (i.e., not too 
different from a Salpeter IMF) and initial cluster structure (basically, any 
initial density profile that is ``not too centrally concentrated'').
If we assume that core collapse corresponds to the onset of runaway collisions,
then the condition for a runaway to occur can be written very simply
$$ \tcc \simeq 0.1\, \trh < \tstar(m_{\rm max}), $$
where $\tstar(m_{\rm max})$ is the lifetime of the most massive stars, of mass 
$m_{\rm max}$.  From current stellar evolution calculations 
\citep[e.g.,][]{SSMM92} we know that $$ \tstar(m_{\rm max}) 
\simeq 3\,{\rm  Myr}~~~~~~{\rm for}~~m_{\rm max}> 30\,M_\odot,$$ i.e., nearly 
{\sl constant\/} (this is simply because these massive stars are nearly
Eddington-limited\footnote{If the mass segregation time scale $t_{\rm s}\propto 
1/m$ is compared to the ``usual'' $\tstar\propto 1/m^3$, one could conclude 
erroneously that massive stars never play a role in core collapse 
\citep{Applegate86}. However, the approximate $1/m^3$ scaling of the stellar 
lifetime applies only for $m\lesssim 10\,M_\odot$.}), and also nearly 
independent of metallicity and rotation (within $\sim 10\%$).  Therefore, 
under very general conditions, the simple criterion for a runaway process 
can be written $$ \trh \lesssim 3\times10^7\,{\rm yr}. $$ This is in perfect 
agreement with the results of the direct $N$-body simulations by 
\citet{PZMcM02}, although they cannot resolve the core collapse in their 
simulations, and instead {\em define\/} core collapse to be the onset of 
collisions\footnote{For any runaway collision process to occur, 
the IMF must extend to $m_{\rm max} \gg 10\,M_\odot$ and the total number
of stars must be sufficiently large, $N \gg 10^3$. Otherwise there will simply 
not be enough (or any) massive stars to collide.}.

Perhaps the most interesting result from our new simulations is that the 
central (BH) mass produced by this runaway process may well be determined 
largely by the Spitzer instability: the total mass in massive stars going into 
core collapse will be the final BH mass, at least in the absence of 
significant mass loss from stellar evolution. Remarkably, we find that this 
mass is always (within the same ``reasonable'' assumptions about the IMF and 
initial cluster structure as above) around $10^{-3}$ of the total cluster
mass, as suggested by observations (see Kormendy, Richstone, and van der 
Marel, this volume).

\section{Monte Carlo Simulations of Dense Star Cluster Dynamics}

\def\MCglob{\texttt{MCglob}}
\def\MCnucl{\texttt{MCnucl}}

% Order of MCnucl and MCglob was switched to reflect order of corresponding subsections - F. Rasio
Mass segregation and collisional runaways are driven by the most
massive stars in a cluster, which, for a ``normal'' IMF, represent a
very small fraction of the stellar population. For
instance, in a Salpeter IMF from 0.2 to 120\,$M_\sun$, only a fraction
$\simeq 3\times 10^{-4}$ of stars are more massive than
60\,$M_\sun$. Resolving these processes thus requires numerical simulations
with very large numbers of particles (at least $\sim 10^5 - 10^6$). In the last 
few years, new MC codes have
been developed that make possible simulations of stellar clusters with such
high resolutions. The results reported here have been obtained with
two independent MC codes that we call, for convenience, {\MCglob} and
{\MCnucl}, as they were devised to follow the evolution of globular clusters and
galactic nuclei, respectively. Both are based on the
scheme first proposed by \citet{Henon73}, and they rely on very 
similar principles that we summarize below. In
Sections~\ref{subsec:MCglob} and \ref{subsec:MCnucl}, a few important aspects
specific to each code are described.

The MC technique assumes that the cluster is spherically symmetric and
can be modeled by a set of discrete particles. Each particle in the simulation
could represent an individual star (as in \MCglob), or an entire spherical 
shell of stars sharing the same orbital and stellar properties (as in \MCnucl\ 
and most earlier MC codes). In the latter case, the number of particles may be 
lower than the number of stars in the
simulated cluster, but the number of stars per particle has to be the
same for all particles. Another important assumption in these codes is that the
system is always in dynamical equilibrium, so that orbital time scales
need not be resolved and the natural time step is a fraction of the
relaxation (or collision) time. Instead of being determined by
integration of its orbit, the position $R$ of a particle (star, or the radius
$R$ of the shell) is picked at random, with a probability density that
reflects the time spent at $R$: $\mathrm{d}P/\mathrm{d}R\propto
1/V_\mathrm{r}(R)$ where $V_\mathrm{r}$ is the radial velocity along the orbit.

The relaxation is treated as a diffusive process in the usual FP
approximation \citep{Chandrasekhar60,BT87}.  The long-term
effects on orbits of departures of the gravitational forces from a
smooth quasi-stationary potential, $\phi_\mathrm{s}$, are assumed to
be those of a large number of uncorrelated small-angle scatterings.
If a particle of mass $M_1$ travels with relative
velocity $\upsilon_\mathrm{rel}$ through a homogeneous field of particles of
mass $M_2$ with number density $n$ during $\delta t$, in the
center-of-mass reference frame, its trajectory will be deflected by an
angle $\theta_{\delta t}$ with 
$$
\langle \theta_{\delta t} \rangle = 0 \mbox{\ \ \ \ and\ \ }
\langle \theta^2_{\delta t}
\rangle = 8\pi \ln\Lambda \, G^2 n \left(M_1+M_2\right)^2 \upsilon_\mathrm{rel}^{-3}\,\delta t,
$$
where $G$ is the gravitational constant and $\ln\Lambda\simeq 10-15$ is the 
familiar Coulomb logarithm \citep{BT87}. In the MC codes, at each time step, a
pair of neighboring particles are selected and their velocities are
modified through an effective hyperbolic encounter with deflection
angle $\theta_\mathrm{eff}=\sqrt{\langle \theta^2_{\delta t}
\rangle}$. As any given particle will be selected many times, at various
positions on its orbit, the MC scheme will actually integrate the
effect of relaxation over the particle's orbit and over all possible
field particles. Proper averaging is ensured if the time steps are
sufficiently short for the orbit to be significantly modified only
after a large number of effective encounters.  The gravitational potential 
$\phi_\mathrm{s}$ is approximated as the potential generated by all
the particles. This potential is not completely smooth because the
particles are razor-thin spherical shells whose radii change
discontinuously. Through test computations, it can be shown that the
corresponding spurious relaxation is negligible if the number of
particles is $\gtrsim 10^4$ \citep{FB01a}.

In contrast to methods based on the direct integration of the 
FP equation in phase space, the particle-based MC approach allows for the 
natural inclusion of many additional stellar and dynamical processes, such 
as stellar evolution, collisions and mergers, tidal disruptions, captures, 
large-angle scatterings, and strong interactions with binaries. 
The dynamical effects of binaries (the dominant 3- and 4-body
processes), which may be crucial in the evolution of globular
clusters, have been included in several MC codes through the use of
approximate analytic cross sections and simple recipes
\citep{Stodol86,GS00,FGJR03}. Codes are currently in development by
Giersz, Rasio and collaborators that will treat
binaries with much higher realism by explicitly integrating 3- or
4-body interactions on the fly, a ``brute-force'' approach required to
tackle the full diversity of unequal-mass binary interactions.

Among methods that can be used to follow the dynamical evolution
of collisional stellar systems, only direct $N$-body integrations do not
require assumptions on the geometry of the system or dynamical
equilibrium and, being also particle based, rival the ability of MC
codes to incorporate realistic physics. Unfortunately, as they are
based on explicit integration of the orbits of $N$ particles and
require the computation of all 2-body forces, they are extremely
computationally demanding, with a CPU time scaling like $N^{2-3}$. In
practice, even with the use of special-purpose computers (such as the
current GRAPE-6) and in spite of continuous progress in the development of 
$N$-body algorithms, the simulation of a cluster containing $\sim 10^5$ stars 
still requires months of computer time \citep{Makino01}. In contrast, MC codes, 
with CPU times scaling like $N\ln(N)$, routinely use up to a few million 
particles. Such high numbers of particles imply that globular clusters can 
actually be modeled on a star-by-star basis
\citep{Giersz98,Giersz01,JRPZ00,JNR01,WJR00}. However, 
galactic nuclei typically contain $N_\ast \approx 10^7-10^8$ stars,
and, for such systems, one has to take advantage of the fact that each
physical process is included in the simulation with its explicit
$N_\ast$ scaling so that a single particle can represent many 
stars (in \MCnucl). 

\subsection{A Monte Carlo Code for Globular Cluster Simulations}
\label{subsec:MCglob}

One of our MC codes, {\MCglob}, based directly on the ideas of 
\citet{Henon73} described in the previous section, was developed to study
the dynamical evolution of globular clusters using a realistic number of 
stars \citep{JRPZ00,JNR01,FGJR03}. One important characteristic of this 
code is that each particle represents a single star (or binary star) in the 
cluster, which allows us to incorporate stellar evolution processes in a 
completely realistic manner. The code uses a single time step for the evolution 
of all the stars (typically a small fraction of the core relaxation time), 
which allows for effective parallelization of the algorithm.  Typical 
simulations for $\sim 10^5$ stars over $\sim 10^{10}\,$yr can be performed 
in just a few hours of computing time.

This code has been tested extensively using comparisons to previous FP 
and direct $N$-body results, and it has been used to study a variety of 
fundamental dynamical processes such as the Spitzer instability \citep{WJR00} 
and mass segregation \citep{FJPZR02} in simple two-component clusters. It was 
also used to re-examine the question of globular cluster lifetimes 
in the tidal field of a galaxy \citep{JNR01}.  Work is in progress to 
incorporate a full treatment of primordial binaries, including all dynamical 
interactions (binary--binary and binary--single) as well as binary stellar 
evolution \citep{RFJ01,FGJR03}.  Another major advantage of these star-by-star 
MC simulations (e.g., compared to direct FP schemes) is that it is 
straightforward to include a realistic, continuous stellar-mass spectrum. 
With a broad IMF, this requires adjusting carefully the Coulomb logarithm, 
which can be done by using the results of large $N$-body simulations for 
calibration \citep{GH96,GH97}.  Another difficulty introduced by a broad 
%XX missing GH96, GH97 % Corrected (Marc Freitag)
mass spectrum is the necessity of adjusting carefully the time step to treat 
correctly encounters between stars of very different masses. When pairs of 
stars are selected to undergo an effective hyperbolic encounter as described 
above, one has to make sure that the deflection angle remains small for 
{\em both\/} stars. In situations where the mass ratio of the pair 
can be extreme, one has to decrease the time step accordingly \citep{Stodol82}.
%XX missing Stodel82 % Corrected (Marc Freitag)
In practice, for the simulations described here, we find that the time step has 
to be reduced by a factor of up to $\sim10^3$ compared to what would be 
appropriate for a cluster of equal-mass stars.

\subsection{A Monte Carlo Code for Galactic Nuclei Simulations}
\label{subsec:MCnucl}

To the best of our knowledge, {\MCnucl} is the only H\'enon-like MC code 
specifically aimed at the study of galactic nuclei 
\citep{Freitag01,FB01a,FB02b}. In addition to relaxation and stellar
evolution, it also incorporates a detailed treatment of collisions between
single main-sequence stars. A central massive BH can also be included 
(together with a full treatment of tidal disruptions and stellar captures) but
this is not necessary for the work presented here.

Individual time steps are used, with particles updated more frequently where 
the evolution is faster. Specifically, the time steps are set to some small 
fraction $f$ of the {\em local\/} relaxation (or collision) time: 
$\delta t(R) \simeq f \left(\trel^{-1} + \tcoll^{-1}\right)^{-1}$. At each
step, a pair of neighboring particles is selected randomly with
probability $P_\mathrm{selec} \propto 1/\delta t(R)$, ensuring that
the {\em average\/} residence time at $R$ is $\delta t(R)$. After a
particle is modified, the potential, stored in a binary-tree
structure, is updated.

Unlike relaxation, collisions cannot be treated as a
continuous process. They are discrete events that can affect very significantly
the orbits and masses, or even the existence of particles. When a
pair is selected, the collision probability between stars from each
particle (shell),
$$
P_\mathrm{coll} = S_\mathrm{coll} \upsilon_\mathrm{rel}
n \, \delta t \mbox{\ \ \ with\ \ }
\nonumber S_\mathrm{coll} = \pi b_\mathrm{max}^2 = \pi (R_1+R_2)^2
\left(1+\frac{2G(M_1+M_2)}{(R_1+R_2)\upsilon_\mathrm{rel}^2}\right),
$$
is compared with a uniform-deviate random number to decide whether a
collision has occurred. If so, the impact parameter $b$ is determined by
picking another random number $X$, with $b = b_\mathrm{max} \sqrt{X}$. The
other parameters of the collision, i.e., $M_1$, $M_2$ and 
$\upsilon_\mathrm{rel}$, 
are known from the particle properties. The final outcome of the collision 
(new velocities and masses) is determined very accurately by interpolation 
from a table containing the results of more than 14,000 3-D SPH (smoothed 
particle hydrodynamics) calculations of encounters between
two main-sequence stars \citep{TheseFreitag,FB00c,FB03}. 

The structure and stellar evolution of collision products is an
intricate problem that has only been studied in some detail for
low-velocity collisions, relevant to globular clusters
\citep{SLBDRS97,SFLRW00,LWRSW02}. The main issues are the importance of
entropy stratification versus collisional mixing, and the rapid rotation 
of merger remnants. In view of these difficulties, we use one of two simple 
prescriptions to treat the stellar evolution of collision products. {\em (1) 
Maximal rejuvenation,} where the remnant is assumed to be completely mixed and
is brought back on the zero-age main sequence. As hydrodynamic simulations show
only very little mixing, this assumption leads obviously to an overestimate
of the stellar lifetime of collision products. {\em (2) Minimal 
rejuvenation.}  Here we assume that, during a merger, the helium cores of 
both parent stars merge together, while the hydrogen envelopes combine to 
form the new envelope; no hydrogen is brought to the core. An effective age on
the main sequence is given by adopting a linear dependence of the helium 
core mass on age.

\begin{figure}
%\begin{center}
%\resizebox{\hsize}{!}{\includegraphics{LR_fig1.eps}}
%\end{center}
\plotone{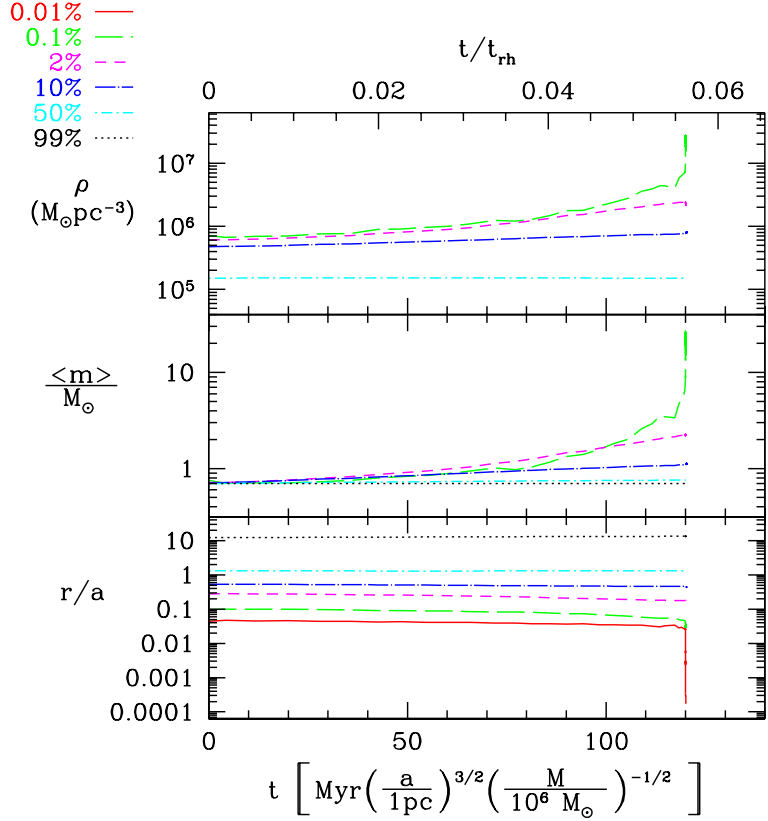}
\caption{\label{LR1}
The evolution of a cluster containing $N = 2.5\times10^6$ stars, terminated 
at core collapse. Time is given both in years (bottom) and in units of the 
initial half-mass relaxation time (top).  The initial configuration is a 
Plummer sphere with a Salpeter IMF spanning the range from 
$m_{\rm min}=0.2\,M_\odot$ to $m_{\rm max}=120\;M_\odot$. The bottom panel 
shows the evolution of Lagrange radii (enclosing a constant fraction of the 
total cluster mass, indicated in the top left), in units of the Plummer length 
(core radius) of the initial model. The middle panel shows the average 
stellar mass within each Lagrange radius, and the top panel shows the average 
density within each Lagrange radius. Core collapse takes place at 
$t = 0.056\, t_{\rm rh}(0)$.}
\end{figure}

\begin{figure}
\begin{center}  
\resizebox{\hsize}{!}{\includegraphics{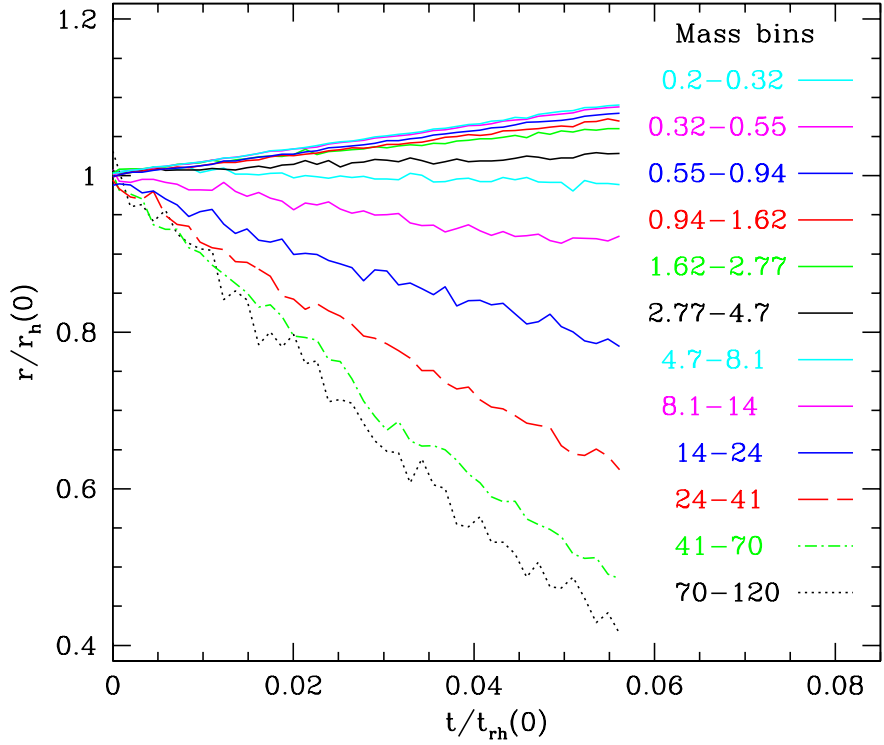}}
\end{center}
%\plotone{mass_bins.eps}
\caption{\label{mass_bins}
The evolution of the mean radius (in units of the initial half-mass radius)
for stars in various mass bins (indicated on the right; values are in units of 
$M_\odot$).  The simulation is the same as in Fig.~\ref{LR1}.  Mass 
segregation is very clear, concentrating more massive stars near the center 
of the cluster.}
\end{figure}

\section{Core Collapse in Young Star Clusters}

\begin{figure}
%\begin{center}
%\resizebox{\hsize}{!}{\includegraphics{LR_fig2.eps}}
%\end{center}
\plotone{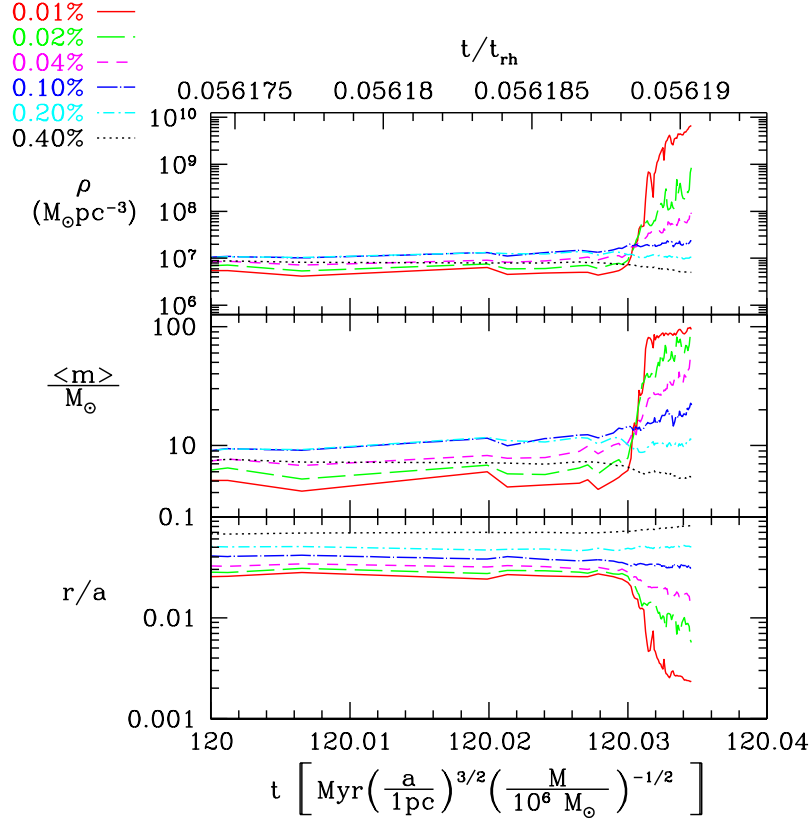}
\caption{\label{LR2}
Same as figure \ref{LR1}, but concentrating on the evolution of the cluster
core near collapse. }
\end{figure}

We have used our code {\MCglob} to study the onset of core collapse in young 
star clusters with a wide variety of initial structures and stellar IMFs 
\citep{GFR03}.  The results of a typical simulation are illustrated in 
Figs.~\ref{LR1}--\ref{LR2}.  This simulation was performed for a cluster 
containing $2.5\times10^6$ stars with a Salpeter IMF between 
$m_{\rm min}=0.2\,M_{\odot}$ and $m_{\rm max}=120\,M_{\odot}$. The rapid mass 
segregation of the most massive stars toward the cluster center is evident, 
as is the abrupt onset of core collapse once the central region becomes 
dominated by massive stars.  In Fig.~\ref{mass_bins} we show the evolution 
of the mean radius of stars in various mass bins. We see that the mass 
segregation starts right at the beginning of the evolution and proceeds at a 
steady rate, even though the overall mass distribution (total density profile) 
in the cluster is hardly changing. Significant changes in the Lagrange radii 
become apparent only when the heaviest stars reach the center of the cluster. 
This is because these heaviest stars account for only a very small fraction 
of the total cluster mass.  In all our simulations we find that the ratio of 
the core collapse time to initial half-mass relaxation time, 
$t_{\rm cc}/t_{\rm rh}(0)$, is always within the range 0.05--0.20 so long as 
the heaviest stars have masses  $\gtrsim 20$ times the average stellar mass 
in the cluster. We used this result [$t_{\rm cc}/t_{\rm rh}(0)\simeq 0.1$] 
in Section 1.1 to derive our simple criterion for the onset of a runaway.
In Fig.~\ref{LR2}, we show the core collapse in more detail.   We see that 
massive stars representing about $0.1$\% of the total cluster mass are driving 
the final core collapse. This ratio of the collapsing core mass to total 
cluster mass, $M_{\rm cc}/M_{\rm tot}$, also appears to be confined to 
a narrow range, of about 0.1\%--0.3\%, in all our calculations to date. These 
were performed for both Plummer models and King models with varying initial 
concentration, and for a variety of IMFs including simple power-law IMFs
with exponents in the range 2--3 (the standard Salpeter value being 2.35). 
We have also checked the robustness of this result by varying the number of 
stars and the time step in our simulations. As pointed out in Section 1.1, the 
apparent agreement between this ratio and the ratio of BH mass to total 
cluster mass in many observed systems is very encouraging.  Work is in 
progress to study more systematically the dependence on initial cluster 
structure, including the possibility of initial mass segregation.

\section{The Runaway Collision Process}
\label{sec:runaway}

\begin{figure}
%\begin{center}
%\resizebox{\hsize}{!}{\includegraphics{iso_trel_Plummer.eps}}
%\end{center}
\plotone{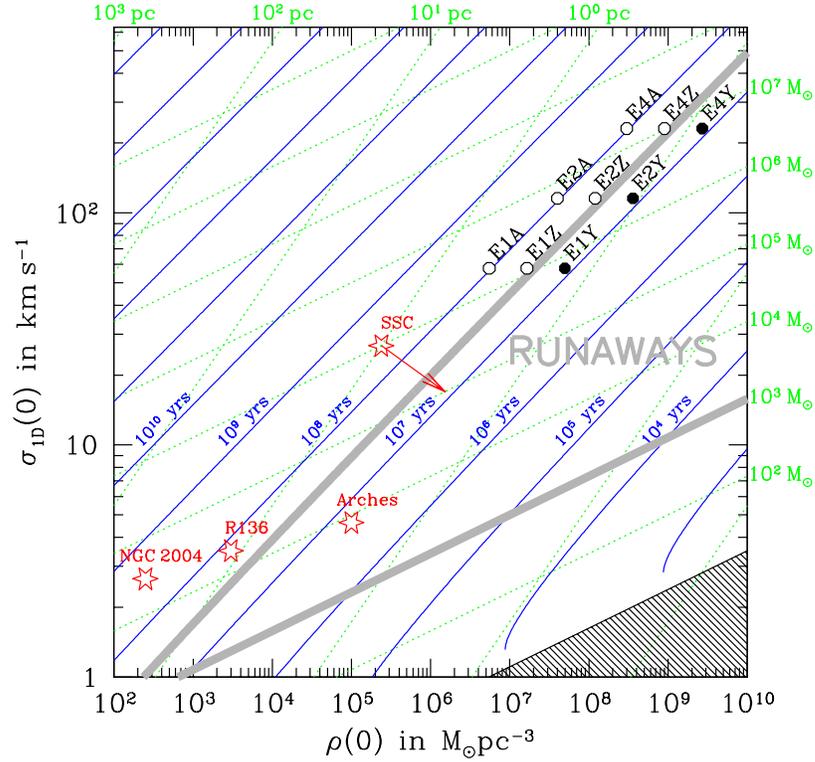}
\caption{Initial conditions for the clusters considered in
Sec.~\ref{sec:runaway}.  We use a notation similar to that of QS90, but our 
models have a broad IMF.  Here $\rho(0)$ and $\sigma_\mathrm{1D}(0)$ are 
the initial central density and velocity dispersion, respectively. The thin 
dotted lines show the values of the total mass (right labels) and the 
initial Plummer scale (top labels). We also show lines of constant
half-mass relaxation time (solid lines labeled from $10^4$ to $10^{10}$ yrs).  
Clusters born in the region between the thick diagonal lines have a core 
collapse time shorter than the lifetime of their most massive stars and 
are expected to undergo a runaway collision process. The solid and open round 
%XXX where are the open dots?
dots in the upper right region show the results of our MC simulations: an 
open dot indicates that a runaway was avoided, while a solid dot indicates 
that the onset of runaway collisions was detected.  The open star symbols 
indicate the positions of a few observed young star clusters. The one labeled 
``SSC'' indicates the position of a typical ``super star cluster.'' Most of 
these systems have sizes at the resolution limit of the observations, with an 
upper limit on their radius of $\sim 1\,$pc.}
\label{fig:condini}
\end{figure}

\begin{figure}
%\begin{center}
%\resizebox{\hsize}{!}{\includegraphics{evol_raylag_PC_1364.eps}}
%\end{center}
\plotone{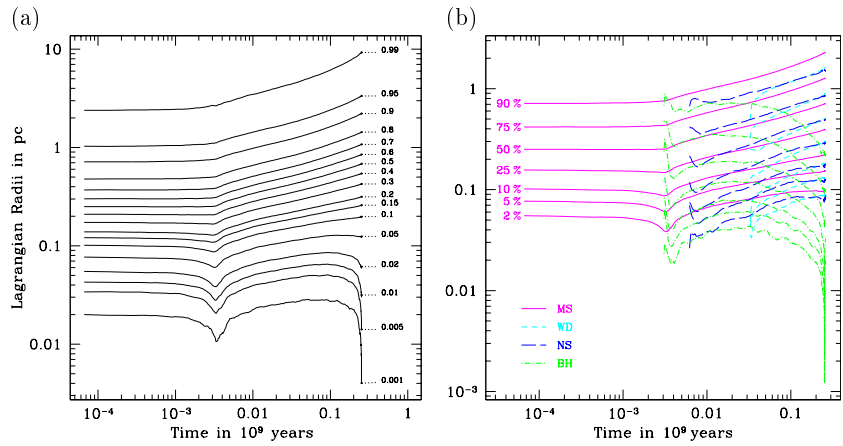}
\caption{Evolution of a cluster model with an extended IMF and an
  initial density 3 times higher than model E4A but the same
  velocity dispersion (see text). {\em (a)} Evolution of the overall
  Lagrange radii. The first, very mild core
  collapse is driven by the segregation of the massive stars to the
  center. It is quickly reversed by their evolution and the associated mass
  loss. The second, much deeper core collapse is driven by the stellar-mass
  BHs, which have become the most massive species after a few Myrs
  ($7\,M_\odot$). {\em (b)} Lagrange radii for the various
  stellar species.}

\label{fig:E4Z}
\end{figure}

\begin{figure}
%\begin{center}
%\resizebox{0.9\hsize}{!}{\includegraphics[bb=81 160 493 707,clip]{MostCollidPart_PC_1361.eps}}
%\end{center}
\plotone{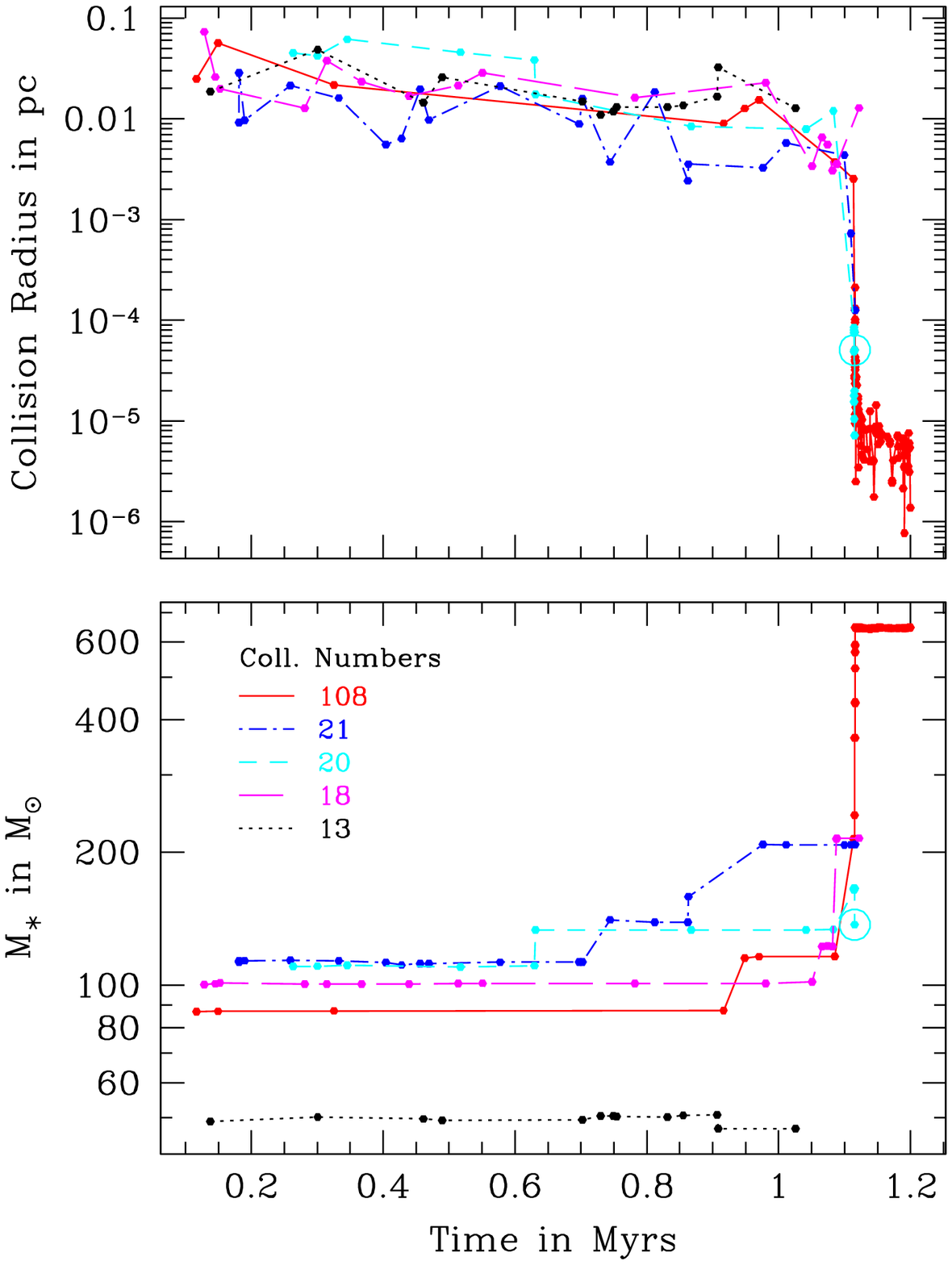}
\caption{Histories of a few particles that suffered from a large number of 
collisions. The initial model, E4Y, has an extended IMF and is 9 times 
denser than E4A but with the same velocity dispersion (see text). The top 
panel shows the distance from the center at which each collision occurred. In 
the bottom panel we plot the mass of the star after each collision. One 
particle (solid line) undergoes runaway growth up to about $650\,M_\odot$. The
reason why the growth saturates abruptly at this value is still unclear to us.
The circle indicates that the particle has merged with a more massive one, 
probably the runaway star.}

\label{fig:runaway}
\end{figure}

The formation of a massive central object by runaway collisions and mergers
has been demonstrated in idealized FP models of proto-galactic nuclei by QS90.
Unfortunately, because of the limitations of 
FP codes, these authors had to rely on a highly simplified
treatment of collisions (see below). Furthermore, they assumed an
initial single-mass stellar population, which significantly reduces
the impact of mass segregation and stellar evolution. Indeed, these
processes come into play only when more massive stars are formed
through collisions.

Using {\MCnucl}, we have started re-examining this problem with more
realistic simulations. In the high-velocity environment of galactic
nuclei, the formation and/or survival of binaries is unlikely (i.e.,
most binaries are ``soft''), so it is reasonable to neglect them in the
computations. As the stellar density rises abruptly to very high values
during core collapse, collisions are bound to
occur even in the absence of binaries. 

For definiteness, we consider first QS90's model E4A, starting from a Plummer 
sphere with initial central density and 3D velocity
dispersion of $3\times 10^8\,M_\odot\mathrm{pc}^{-3}$ and
$400\,\mathrm{km}\,\mathrm{s}^{-1}$. QS90 started their FP simulations
with all stars having $1\,M_\odot$ and assumed that all collisions
lead to mergers with no mass loss and maximal rejuvenation.  Not
surprisingly, if we use the same, highly simplified treatment of
collisions as QS90, we get clear runaway growth of one or a few
particles. When we switch to our realistic prescription for the
collisions and minimal rejuvenation, the runaway still occurs, although
later. However, if we adopt a more realistic, Kroupa-type IMF
\citep{Kroupa00b}, significant mass loss from the massive stars occurs
before core collapse has proceeded to high stellar densities. As we
assume that the gas is not retained in the cluster, this mass loss
drives a re-expansion of the whole system. A second, deeper core
collapse occurs later, when the remnant (stellar-mass) BHs segregate 
to the center. The evolution of this dense cluster of stellar BHs
cannot be treated with the present version of {\MCnucl} because
dynamically formed binaries will play a central role.

In addition to models with the same densities and velocity dispersions
as the class ``A'' considered by QS90, we also simulated clusters with
densities 3~times (models ``Z'') and 9~times (models ``Y'') larger (but the same
velocity dispersion) with correspondingly shorter relaxation times
(see Fig.~\ref{fig:condini}). As shown in Fig.~\ref{fig:E4Z},
models of class ``Z'' have a core collapse time slightly larger
than 3\,Myr and do not enter a runaway phase.  One the other hand, runaway 
growth occurs in all simulations for clusters of class ``Y,'' which collapse in 
about 1\,Myr. Figure~\ref{fig:runaway} shows such a case. The growth of the
runaway particle(s) is limited to a few hundred $M_\odot$
($650\,M_\odot$ in the ``best'' case), probably by some numerical artifact.  
Note that 500,000 particles were used for all these computations, independent 
of the true number of stars in the cluster. Hence, every particle represents 
many stars (12 to 36), a numerical treatment whose validity becomes obviously 
questionable as soon as a single particle detaches from the overall mass
spectrum. In addition, we note that, before the runaway abruptly saturates 
in our simulations, the growth rate observed is extremely rapid and the basic 
orbit-averaging assumption implied in the MC technique must probably break 
down. Possible physical processes that could terminate the runaway include:
rapid mass loss from some of the massive stars, increased inefficiency of
collisional merging for very massive objects\footnote{We have not computed 
collisions for stars more massive than $75\,M_\odot$, so considerable
extrapolation of our results is required in these simulations.}, depletion of 
the ``loss-cone'' orbits that bring stars to the center, or some
combination of these factors. In any case, a robust conclusion can be
drawn from these simulations, namely that runaway merging can produce
stars at least as massive as $\sim 500\,M_\odot$ in the centers of clusters
with $t_{\rm cc} \lesssim 3\,\mathrm{Myr}$, even when central velocity 
dispersions are as high as $\sim 400\,\kms$.

\section*{Acknowledgements}

% Slight changes made here.  -F. Rasio
We are very grateful to Douglas Heggie, Steve McMillan, and Simon Portegies 
Zwart for many helpful discussions. This work was supported by NASA ATP Grant 
NAG5-12044, NSF Grant AST-0206276, and NCSA Grant AST980014N at Northwestern University. 
% Modified by Marc 03/28/03 \\\
The work of MF was supported in part by Sonderforschungsbereich (SFB)
439 'Galaxies in the Young Universe' of the German Science Foundation
(DFG) at the University of Heidelberg, performed under subproject A5.
% End modification ///

%\bibliographystyle{apj} 
%
%\bibliography{aamnem99,biblio} 

\end{document}